\documentstyle[11pt]{article}
\def\beq{\begin{equation}}
\def\eeq{\end{equation}}
\def\beql{\arraycolsep .5mm \begin{eqnarray}}
\def\eeql{\end{eqnarray}}
\def\zeile{\nonumber \\[2mm] }
\def\zeilex{ \\[2mm] }
\def\rf#1{(\ref{#1})}
\def\d{{\rm d}}
\def\x{{\bf x}}

\def\eps{\varepsilon}
\def\Tr{{\rm Tr}}
\def\det{{\rm det}}
\def\del{\partial}
\def\eins{\mbox{\bf \sf 1}}
\def\ft#1#2{{\textstyle{{#1}\over{#2}}}}
\def\pois#1#2{ \{ #1 , #2  \}}
\def\te{\tilde e}
\def\dotte{\dot{\tilde e}}
\def\f#1{\varphi^{#1}}
\def\dotf#1{\dot \varphi^{#1}}
\def\p#1{\pi^{#1}}

\def\H{H}
\def\C{C}

\def\imat{I}
\def\J{J}
\def\jmat{M}

\def\Pexp{{\c P} \, \exp\, }
\def\c#1{{\cal #1}}
\def\oper#1{\widehat{#1}}
\def\deltadelta#1/#2/{\frac{\delta #1}{\delta #2}}

\def\SO{{\rm SO}}
\def\SU{{\rm SU}}
\def\U {{\rm U }}

\begin{document}
gr-qc/9305025
\vspace*{5mm}
\begin{center}
\LARGE
Solutions to the Wheeler DeWitt Constraint of Canonical
Gravity Coupled to Scalar Matter Fields\\
\vspace*{1cm}
\normalsize
Hans-J\"urgen Matschull \\
II. Institut f\"ur theoretische Physik\\
Universit\"at Hamburg \\
Luruper Chaussee 149\\
W-2000 Hamburg 50 \\
Germany\\
May 26, 1993\\
\vspace*{2cm}

\begin{minipage}{11cm}
\small
It is shown that the Wheeler DeWitt constraint of canonical gravity
coupled to Klein Gordon scalar fields and expressed in terms
of Ashtekar's variables admits formal solutions which are parametrized
by loops in the three dimensional hypersurface and which are
extensions of  the well known Wilson loop solutions found by
Jacobson, Rovelli and Smolin.
\end{minipage}
\end{center}
\vspace*{1cm}
\section*{Notation and Constraints}

The action of four dimensional gravity coupled to $N$ real scalar
matter fields is given by
\beq
 \label {action}
    S[E_\mu^A , \f m]    =
        \int \d^4\!x \, \ft12 E \Big(  R
      -   G^{\mu\nu} \, \del_\mu \f m \del_\mu \f m
      -   m^2 \, \f m \f m \Big) ,
\eeq
where $\mu, \nu, \dots$ denote four dimensional curved indices,
$A,B,\dots$ are flat indices raised or lowered by the
Lorentzian metric with signature $(-,+,+,+)$.
$E$ is the determinant and $G_{\mu\nu}$ the metric generated
by the vierbein $E^A_\mu$.
$R$ is the metric curvature scalar obtained from $G_{\mu\nu}$.
                             Following the procedure in
\cite{arnowitt.deser.misner:62} and
\cite{misner.thorn.wheeler:73}, but with the
vierbein components as the basic variables instead of the metric,
spacetime is split into space and time by decomposing
the vierbein into a dreibein $e_a^\mu$, the lapse function $n$, and
the shift vector $n^\alpha$:
\beq
      E_\mu^A =
                \pmatrix{  n        & 0   \cr
                           n^\beta e_{\beta a}   & e_{\alpha a}   \cr}.
\label {decomposition}
\eeq
Thereby the four dimensional flat index $A$ splits into
the three dimensional index $a$, labeling the three spatial
unit vectors, and the index $0$, indicating the timelike
unit vector. As the three metric is positive definite,
$a,b,\dots$ will always appear as lower indices.
Likewise, the curved index $\mu$ is split into the
tangent space index $\alpha$ of the spatial three manifold and the
index $t$, denoting the (parameter) time coordinate%
\footnote{For a detailed review of the space time split
in terms of the vierbein and with similar conventions
see also
\cite{matschull.nicolai:92}.}.

To treat this theory canonically `a la Dirac'\cite{dirac:65},
one has to express the action as a time integral of the
Lagrangian and identify the canonical variables, which are most
conveniently taken to be $\te^\alpha_a := e e^\alpha_a$,
$n$, $n^\alpha$, and $\f m$, where $n$ and $n^\alpha$
are Lagrange multipliers, whereas for the other fields we
obtain the momenta
\beql
    \p m & = & \deltadelta \c L / \dotf m / =
       e n^{-1} ( \dotf m - n^\alpha \del_\alpha \f m ),\zeile
    p_{\alpha a} & = & \deltadelta \c L / \dotte_a^\alpha / =
       -  e_a^\beta \, k_{\alpha\beta}
\label {momenta}
\eeql
with the extrinsic curvature
\beq
     k_{\alpha \beta} = \ft12  \, n^{-1} (
         \dot g_{\alpha\beta} - \nabla\!_\alpha n_\beta
                              - \nabla\!_\beta n_\alpha ).
\eeq
The basic Poisson brackets are
\beq
   \pois { \te_a^\alpha  } { p_{\beta b} }
       = \delta_\beta^\alpha \delta_{ab} , \ \ \ \ \
   \pois { \f m } { \p n } = \delta^{mn},
\eeq
which, of course, have to be supplemented by spatial delta functions.
The derivatives of the Lagrangian with respect to $n$ and $n^\alpha$
as well as the symmetry of $k_{\alpha\beta}$
yield the Wheeler DeWitt\cite{wheeler:64,dewitt:67},
diffeomorphism, and $\SO(3)$
constraints, respectively.
Expressed in terms of
Ashtekar's variables\cite{ashtekar:8687},
which are defined by
\beq
     A_{\alpha a} := i p_{\alpha a} + \ft12 \eps_{abc}
                        e_a^\beta \nabla\!_\alpha e_{\beta b},
\eeq
instead of the momentum variables $p_{\alpha a}$,
the constraints read
\beql
    \H & = &  \ft12 \eps_{abc} \, \te_a^\alpha \te_b^\beta \,
           F_{\alpha\beta c}
        - \ft 12 \, \p m \p m \zeile & &  {}
  - \ft12 \,\te\,  m^2 \, \f m \f m
 - \ft12  \, \te_a^\alpha  \del_\alpha \f m
           \, \te_b^\beta   \del_\beta \f m,
     \zeile
    \H_{\alpha} & = &  - i \, \te_a^\beta \, F_{\alpha \beta a}
                - \del_\alpha \f m \p m , \zeile
      \C_a & = & D_\alpha \te_a^\alpha ,
\label {const}
\eeql
where $\te=e^2$ is the determinant of $\te_a^\alpha$
and      $A_{\alpha a}$ is interpreted as a $\SO(3)$ connection
defining a covariant derivative
$D_\alpha V_a := \del_\alpha V_a + \eps_{abc}
A_{\alpha b} V_c$ and with curvature
$F_{\alpha\beta a}= \del_\alpha A_{\beta a}
- \del_\beta A_{\alpha a} + \eps_{abc} A_{\alpha b} A_{\beta c}$.

The non vanishing Poisson brackets are now given by
\beq
     \pois { \te^\alpha_a  } { A_{\beta b} }
      =  i \delta_\beta^\alpha \delta_{ab} , \ \ \ \ \
   \pois { \f m } { \p n } = \delta^{mn}.
\label {A-pois}
\eeq

\section*{Quantization and Observables}

The quantum operators for the canonical variables are
obtained by requiring the commutators to be $-i$ times
the Poisson brackets. The wave functional $\Psi$ will be
a function of $A_{\alpha a}$ and $\f m$, and the
other fields are represented by
\beq
 \oper{\te_a^\alpha}  =  \deltadelta / A_{\alpha a} / , \ \ \ \ \
 \oper{\p m}  =  i \deltadelta / \f m /  .
\label {op-rep}
\eeq

With this representation chosen and with the ordering
such that all differential operators appear to the right,
the $\SO(3)$ and diffeomorphism constraints in fact generate
$\SO(3)$ rotations and diffeomorphisms on the wave functional,
and thus they require it to be invariant under these transformations.

The Wheeler DeWitt constraint is represented by the operator
\beql
   \oper \H   =
       \ft12 \big( \eps_{abc} F_{\alpha\beta c}
       -  \delta_{ab} \del_\alpha \f m   \del_\beta \f m \big)
       \,   \deltadelta / A_{\alpha a} /
          \deltadelta / A_{\beta b } / && \zeile
        {} - \ft1{12} \eps_{abc} \eps_{\alpha\beta\gamma}
      \,  m^2\, \f m \f m
       \,   \deltadelta / A_{\alpha a} /
       \,   \deltadelta / A_{\beta  b} /
          \deltadelta / A_{\gamma c} / && \zeile
                    {}  + \ft 12 \,  \deltadelta / \f m /
                    \deltadelta / \f m /, &&
\label {q-wdw}
\eeql
which, of course,  needs a regularization.
This problem is discussed in detail in
\cite{jacobson.smolin:88,rovelli.smolin:90}, where one concludes
that a the loop representaion is needed to provide a rigorous
regularization such that the solutions are annihilated by the
constraints in the no-regularization limes. So let us first
obtain the results formally.

In contrast to pure gravity, for the matter coupled model
there are well defined
observables in the sense of Dirac\cite{dirac:65}, i.e.
operators commuting weakly with all the constraints.
They are
the conserved `angular momenta' of the matter
fields, i.e. the Noether charges of the global $\SO(N)$
invariance of the Lagrangian:
\beq
    \J^{mn} := \int \d\x\,  \f {[m} \p {n]} .
\eeq
The corresponding quantum operator is
\beq
  \oper \J(\jmat) :=  i \int \d\x\, \f m \jmat^{mn}
       \, \deltadelta / \f n / ,
\label {observables}
\eeq
for any antisymmetric matrix $\jmat$.

\section*{Solutions}

The well known solutions of the Hamiltonian constraint
for pure gravity parametrized by loops $\gamma$
are constructed out of the Wilson loop integral%
\cite{jacobson.smolin:88,rovelli.smolin:90}
\beq
   T_\gamma(a,b) := \Pexp \ft i2
             \int_a^b \d s \, \gamma^{\prime\alpha}
    \,  A_{\alpha a}\sigma_a,
\label {wilson}
\eeq
where path ordering is defined such that factors nearer to the
upper border $b$ appear to the right, and $\sigma_a$
are the Pauli matrices obeying $\sigma_a \sigma_b = \delta_{ab}\eins
+ i \eps_{abc} \sigma_c$. The prime denotes the derivative with
respect to the curve parameter $s$. To obtain
the action of the Hamiltonian constraint on
$T_\gamma(a,b)$, one has to use
\beql
   \deltadelta / A_{\alpha a}(\x) /
   \deltadelta / A_{\beta  b}(\x) / && T_\gamma(a,b) =  \zeile
        -\ft12
     \int_a^b \d s \int_s^b  \d t &&
         \delta(\x,\gamma(s)) \,  \delta(\x,\gamma(t)) \,
         \gamma^{\prime\alpha}(s)\,  \gamma^{\prime\beta}(t) \,
 \zeile &&{} \
       T_\gamma(a,s) \sigma_a T_\gamma(s,t) \sigma_b T_\gamma(s,b).
\label {ddA}
\eeql
As there is a contribution to the integral only for $s=t$, if the
loop has no self intersections, the expression \rf{ddA} becomes
symmetric in $\alpha,\beta$ and we may assume that
$T_\gamma(s,t)=\eins$. Defining the trace of the holonomy of
the closed loop $\gamma$ by $T_\gamma := \Tr \, T_\gamma(a,b)$,
where $\gamma(a)=\gamma(b)$, we have
\beq
   \oper \H \, T_\gamma =
        \ft38\,  T_\gamma  \,
     \oint \d s \oint \d t \,
         \delta(\x,\gamma(s)) \,  \delta(\x,\gamma(t)) \,
         \del_s \f m          \del_t \f m     .
\label {HT}
\eeq

This expression has to be canceled against an equal
expression that results form the momentum part of the
Wheeler DeWitt constraint acting on a functional that depends on the
matter fields.
Define
\beq
   L_\gamma^\imat := \ft{\sqrt 3}4 \,
                \oint_\gamma \d s\,  \f m  \del_s \f n \, \imat^{mn},
\label {Ldef}
\eeq
where $\imat$ is a real antisymmetric matrix.
Then
\beql
\label {HL}
   \oper \H \exp(i L_\gamma^\imat) =\ \ && \zeilex
  \ft38\,  \exp(iL_\gamma^\imat) &&
   \oint \d s \oint \d t \,
         \delta(\x,\gamma(s)) \,  \delta(\x,\gamma(t)) \,
         \del_s \f m \imat^{mn} \imat^{np}  \del_t \f p     .
\nonumber
\eeql
Obviously, the wave functional
\beq
   \Psi_\gamma^\imat := T_\gamma \, \exp(i L_\gamma^\imat)
\label {sol1}
\eeq
is a solution to the Wheeler DeWitt constraint \rf{q-wdw}, if
$\imat$ is a complex unit, i.e. $\imat^2=-\eins$, as the various
parts of \rf{q-wdw} act either only on $T_\gamma$ or
on $L_\gamma^\imat$.

Further solutions can be obtained by acting on the
functional with the observables \rf{observables}:
\beq
           \oper \J(\jmat)   \, \Psi_\gamma^\imat
  = T_\gamma \, \exp(i L_\gamma^\imat) \, L_\gamma^{[\imat,\jmat]}
\label {sol2}
\eeq
is a new solution to \rf{q-wdw}.

We get nonvanishing state functionals with any
antisymmetric matrix that does not commute with $\imat$.
Those matrices are just the generators of $\SO(N)$ which are
not generators of the $\U(1)\times \SU(N/2)$ subgroup
defined by the complex unit $\imat$.
This is a feature of the matter coupled theory which is
not present in pure gravity. See also \cite{matschull.nicolai:92}
for similar observables in other kinds of matter coupled theories,
e.g. extended supergravity and sigma models.

On the other hand, the main problems of the
loop solutions remain: first of all, the metric determinant
annihilate all the states, and, as a consequence, the states
do not depend on the mass of
the Klein Gordon field. Unfortunately, also the observables
are mass independent      and thus do not produce mass dependent
states.

        One might interpret this on a semiclassical
level by saying that the functionals represent states of an empty
universe, i.e. with no matter present, because only then a
classical solution can be mass independent.
But this is in contradiction with the fact, that there is an
observable representing a classical constant of motion of the
matter fields, which does not annihilate the state functional.
Thus, in a slightly heuristical sense,
there is matter in the universe.
The reason for this apparent contradiction is that in Ashtekar's
formulation of gravity, even on the classical level, there are
solutions to the constraints \rf{const} with matter fields
present, i.e. $\f m\not\equiv 0$, but which are independent
of $m$, namely for $\te=\det g_{\alpha\beta}=0$, i.e. if the spatial
                                 metric is degenerate.

Another property of these solutions is that they are not
diffeomorphism invariant, but this is exactly the same problem
as for pure gravity and may be dealt with in the same way, i.e. by
defining a loop representation and considering wave functionals
defined on equivalence classes of loops\cite{rovelli.smolin:90}.
Rovelli and Smolin showed that this procedure provides a
naturally regularized Hamiltonian and it should be possible
also for the matter coupled theory to make the formal results
obtained here exact in a regularized framework.

A final question is whether there are similar solutions
for other types of matter coupled theories.
Apart from the somewhat awkward factor of $\sqrt3/4$ in the
definition of the `matter loop' in \rf{Ldef}, the
integrand consists essentially of two parts, which may be
generalized to other models. The matrix $\imat$ is a complex
unit on the field space, and
the one form  $\f m \del_\alpha \f n$ integrated along the curve is the
`lower $\alpha$' component of the conserved current
associated with the $SO(N)$ symmetry of the matter fields.
Both quantities have natural generalizations e.g. for
sigma model type matter fields:
$\imat$ may be replaced by an (almost) complex structure and
the global symmetry is provided by a killing vector field on
the target space.
A straightforward generalization of these solutions to
such a sigma model, however, fails, because the killing vector field
and the complex structure depend on $\f m$ in a
nontrivial way and thus extra contributions to
\rf{HL} appear which are not canceled by the corresponding
`gravitational term' \rf{HT}.


\begin{thebibliography}{10}

\bibitem{arnowitt.deser.misner:62}
R. Arnowitt, S. Deser, and C.W. Misner.
\newblock The dynamics of general relativity.
\newblock In L. Witten, editor, {\it Gravitation:
  An Introduction to Current
  Research}, Wiley, New York, 1962.

\bibitem{misner.thorn.wheeler:73}
C.W. Misner, K.S. Thorn, and J.A. Wheeler.
\newblock {\it Gravitation}.
\newblock Freeman, New York, 1973.

\bibitem{matschull.nicolai:92}
H. Nicolai and H.J. Matschull.
\newblock {\it Aspects of Canonical Gravity and Supergravity}.
\newblock Preprint~DESY 92-099, 1992.
\newblock To appear in: {\it Karpacz proceedings}, 1992.

\bibitem{dirac:65}
P.A.M. Dirac.
\newblock {\it Lectures on Quantum Mechanics}.
\newblock Academic Press, New York, 1965.

\bibitem{wheeler:64}
J.A. Wheeler.
\newblock Geometrodynamics and the issue of the final state.
\newblock In C. DeWitt and B. DeWitt, editors, {\it Relativity,
  Groups and Topology}, Gordon and Breach, New York, 1964.

\bibitem{dewitt:67}
B. DeWitt.
\newblock Quantum theory of gravity I\&II.
\newblock {\it Phys. Rev.}, 160:1113, 1967.
\newblock {\it Phys. Rev.}, 162:1195, 1967.

\bibitem{ashtekar:8687}
A. Ashtekar.
\newblock New variables for classical and quantum gravity.
\newblock {\it Phys. Rev. Lett.}, 57:2244, 1986.
\newblock New hamiltonian formulation of general relativity.
\newblock {\it Phys. Rev.}, D36:1587, 1987.

\bibitem{jacobson.smolin:88}
T. Jacobson and L. Smolin.
\newblock Nonpertubative quantum geometries.
\newblock {\it Nucl. Phys.}, B299:295, 1988.

\bibitem{rovelli.smolin:90}
C. Rovelli and L. Smolin.
\newblock Loop space representation of quantum general relativity.
\newblock {\it Nucl. Phys.}, B331:80, 1990.

\end{thebibliography}
\end{document}